\documentstyle[12pt,epsfig,cite]{article}

\begin{document}

\setlength{\unitlength}{1pt}

\epsfysize3cm
%% \epsfbox{belle.eps}    % BELLE-logo
\begin{flushright}
\vskip -3cm
\noindent
%% \hspace*{3.in}BELLE NOTE 891 \ \ \ \ \ \ \ \\
\vspace*{0.5cm}
%% \hspace*{3.in}First version: 1.00 \ \ \ \ \ \ \ \ \\
%% \hspace*{3.in}March 5, 2006 \ \ \ \ \ \ \ \ \ \ \ \ \ \\
% \hspace*{3.in}Revised version: \ \ \ \ \\
% \hspace*{3.in}December 28, 2003
\end{flushright}                   

\begin{center}
\normalfont
\vspace*{3.0cm}

%\title{
\begin{center}
{\Large \bf Determining {\boldmath $\Delta \Gamma_s / \Gamma_s$}
from {\boldmath ${\cal B} (B_s \to D_s^{(*)+} D_s^{(*)-})$} 
measurements at {\boldmath $e^+ e^-$} colliders running on the
{\boldmath $\Upsilon$(5S)} resonance.
}
\end{center}

% \maketitle

\normalsize
\vspace*{1.5cm}
\baselineskip=.6cm
{ A. Drutskoy, \\
{\it University of Cincinnati, US}}\\
\vspace*{1cm}
%\vspace*{2cm}
\end{center}

\vspace*{1cm}

\normalsize
\begin{abstract}
%{\bf Abstract}

The feasibility of determining $\Delta \Gamma_s / \Gamma_s$
from $B_s \to D_s^{(*)+} D_s^{(*)-}$
decay branching fraction measurements at $e^+ e^-$ colliders 
running on the $\Upsilon$(5S) resonance is discussed.
% The $B_s \to D_s^{(*)+} D_s^{(*)-}$ signal efficiency is obtained 
% from the detailed MC simulation of the Belle detector running
% at the KEKB asymmetric energy $e^+ e^-$ collider.
The expected accuracies of the $B_s \to D_s^{(*)+} D_s^{(*)-}$ branching 
fraction measurements and the $\Delta \Gamma_s / \Gamma_s$
determination with \linebreak 50$\,$fb$^{-1}$ of data at the $\Upsilon$(5S)
are estimated using Monte Carlo simulation.
Comparison of that measurement with a directly measured
$\Delta \Gamma_s / \Gamma_s$ value within and beyond Standard Model
is discussed.
A new method of estimating $\Delta \Gamma_s / \Gamma_s$ is proposed,
based on comparing the $CP$-even and $CP$-odd components
in $B^0 \to J/\Psi K^{*0}$ and $B_s \to J/\Psi \phi$ decays.
A comparison of $B_s$ and $B^0$ lifetime measurements with
theoretical predictions is discussed.

\end{abstract}

%\baselineskip=8truept
%\overfullrule=0truept
%{\leftskip=0.5truein
%\rightskip=0.5truein  
%\small
{\renewcommand{\thefootnote}{\fnsymbol{footnote}}

\newpage

\normalsize

\section{Introduction}

The relation between the $B_s$ lifetime difference and the sum of 
$B_s \to D_s^{(*)+} D_s^{(*)-}$ branching fractions 
has been discussed in several theoretical papers \cite{alek,gross,duna,ferm}.
The $B_s$ lifetime difference
between the short-living and long-living components, $\Delta \Gamma_s$,
originates from the $\Delta \Gamma^{CP}_s$ difference 
in the $CP$ eigenstate decay rates (see ref. \cite{duna} for details):

\begin{equation}
 \Delta \Gamma^{CP}_s = \Gamma (B_s^{even}) - \Gamma (B_s^{odd})
\end{equation}

In the Standard Model (SM) the $CP$ violating phase 
$\beta_s = arg({\rm V}_{ts})$ of the 
Cabibbo-Kobayashi-Maskawa matrix element V$_{ts}$ is small.
Therefore, the mass eigenstates coincide with the $CP$ eigenstates.
% and $\Delta \Gamma^{SM}_s = \Delta \Gamma^{CP}_s$.
Potentially, Beyond Standard Model (BSM) physics processes
could provide a large phase, contributing to $B_s-\bar{B}_s$ mixing
and resulting in a reduction of $\Delta \Gamma_s / \Gamma_s$ 
\cite{gross,duna}:

\begin{equation}
\frac{\Delta \Gamma_s}{\Gamma_s} = \frac{\Delta \Gamma_s^{CP}}{\Gamma_s} \ {\rm cos}\,\phi
\end{equation}
where $\phi = arg (- M_{12} / \Gamma_{12})$ and $M_{12}$ and $\Gamma_{12}$
are the $B_s$ mass matrix and decay matrix elements, respectively.
Within the SM the phase $\phi$ is close to zero and a direct
$\Delta \Gamma_s / \Gamma_s$ measurement should coincide with
$\Delta \Gamma_s^{CP} / \Gamma_s$.
Therefore, SM can be tested by comparing the $\Delta \Gamma_s / \Gamma_s$
value obtained from a direct $B_s$ lifetime measurement and 
$\Delta \Gamma_s^{CP}/ \Gamma_s$,
which is determined from the difference between summed $CP$-even and $CP$-odd
$B_s$ decay branching fractions.
The angle $\phi$ (more precisely, sin $\phi$)
can be measured also from a time-dependent analysis of tagged $B_s$ events, 
similar to $B^0$ time-dependent analyses.

The $B_s \to D_s^{(*)+} D_s^{(*)-}$ decays are expected
to dominate the $CP$-eigenstate decays, because of large
branching fractions, of the order of $(1-2)\%$ for each mode.
Moreover, within the heavy quark limit, the 
$B_s \to D_s^{(*)+} D_s^{(*)-}$ decays
are expected to be predominantly $CP$-even states \cite{duna}.
Other $B_s$ decays with fixed $CP$, like $J/\Psi \eta$,
$\phi \phi$ or $K^+ K^-$ modes, have much smaller branching
fractions of the order of 10$^{-4}$, and would lead to
a few percent correction in $\Delta \Gamma_s$ calculations.

Following this approach, the approximate equality was obtained 
in ref. \cite{duna}:

\begin{equation}
2 {\cal B} (B_s \to D_s^{(*)+} D_s^{(*)-}) \approx \Delta \Gamma^{CP}_s / \Gamma_s
\end{equation}

Generally, Eq.(3) was obtained under two assumptions:
a) that $B_s \to D_s^{*+} D_s^-$ and $B_s \to D_s^{*+} D_s^{*-}$ decays 
are predominantly $CP$-even states (expected in the heavy quark limit) 
and b) that $\Delta \Gamma^{CP}_s$ is saturated by the 
$B_s \to D_s^{(*)+} D_s^{(*)-}$ decays (expected in the Shifman-Voloshin 
limit \cite{svl}).
The last assumption means that the contribution of multi-body $CP$-eigenstate 
decay modes, like $B_s \to D_s^+ D_s^- \pi^0$, to $\Gamma (B_s^{even})$ 
(or maybe to $\Gamma (B_s^{odd})$) is small compared to the dominant
contribution of two-body $B_s \to D_s^{(*)+} D_s^{(*)-}$ decays.
The inaccuracy of these two assumptions is expected to be less than
$10\%$ \cite{duna}.

\section{MC simulation of {\boldmath $B_s \to D_s^{(*)+} D_s^{(*)-}$ decays}}

Exclusive $B_s \to D_s^{(*)+} D_s^{(*)-}$ decays are not yet observed.
Moreover, the $D_s^{*}$ mesons cannot be effectively reconstructed in
Tevatron experiments due to very low photon reconstruction efficiency.
However, $B_s \to D_s^{(*)+} D_s^{(*)-}$ decays can be effectively
measured at $e^+ e^-$ colliders running on the $\Upsilon$(5S) resonance.
The CLEO collaboration has recently published the results
of inclusive and exclusive analyses of 0.42$\,$fb$^{-1}$ of 
data \cite{cleoi,cleoe}
which were collected at the $\Upsilon$(5S) resonance,
and a rate of initially produced $B_s$ mesons per 1\,fb$^{-1}$
was observed to be $(1.0 \pm 0.4) \times 10^5$ \cite{cleoi}.
A similar rate of $(0.99 \pm 0.26) \times 10^5$ $B_s$ mesons
per 1\,fb$^{-1}$ has been obtained 
by the Belle collaboration at $\Upsilon$(5S) engineering runs, where a data 
sample of 1.86\,fb$^{-1}$ was collected \cite{beld}.

We performed a full MC simulation of the Belle detector \cite{belled}
for the $B_s \to D_s^{(*)+} D_s^{(*)-}$ decays and
obtained an efficiency of \mbox{$\sim2.0\times10^{-4}$} for 
$B_s \to D_s^+ D_s^-$ decays, \mbox{$\sim1.0\times10^{-4}$}
for $B_s \to D_s^{*+} D_s^-$ (or $B_s \to D_s^+ D_s^{*-}$) decays, and
\mbox{$\sim0.51\times10^{-4}$} for $B_s \to D_s^{*+} D_s^{*-}$ decays.
Assuming ${\cal B}(B_s \to D_s^+ D_s^-) \approx 1\%$,
${\cal B}(B_s \to D_s^{*+} D_s^-) = {\cal B}(B_s \to D_s^+ D_s^{*-})
\approx 2\%$ and ${\cal B}(B_s \to D_s^{*+} D_s^{*-}) \approx 3\%$, we
expect to reconstruct about one event in each decay mode
with an integrated luminosity of $\sim$5\,fb$^{-1}$.
Finally, a 25$\%$ accuracy of $B_s \to D_s^{(*)+} D_s^{(*)-}$
branching fraction measurement can be obtained with an integrated 
luminosity of 50\,fb$^{-1}$, which can be collected during 6-7 weeks
of data taking by the Belle detector running at the $\Upsilon$(5S)
with the current luminosity of the KEKB accelerator 
$\sim 1.6 \times 10^{34}\,$nb$^{-1}\,$sec$^{-1}\,$\cite{kekb}.  

Additionally, the accuracy of the assumptions used to
obtain Eq.~3 can be checked from $\Upsilon$(5S) data.
The $B_s \to D_s^{*+} D_s^{*-}$ decay angular distributions can
be studied to test the $CP$-even component domination assumption.
At $e^+ e^-$ collider experiments the relative contribution of multi-body 
$B_s \to D_s^+ D_s^- \pi^0$ decays or high-lying $D_{sJ}$ final state
$B_s \to D_s^+ D_{sJ}^-$ decays can be estimated searching directly
for these decay modes or other similar final states.

\section{Experimental data and estimates 
of {\boldmath $\Delta \Gamma_s / \Gamma_s$}}

The direct measurements and indirect estimates of 
$\Delta \Gamma_s / \Gamma_s$ are reviewed in this section.

\subsection{Direct measurements of {\boldmath $\Delta \Gamma_s / \Gamma_s$}}

As noted above, the SM can be tested comparing the sum of 
$B_s \to D_s^{(*)+} D_s^{(*)-}$ branching fractions
with a direct lifetime measurement of $\Delta \Gamma_s / \Gamma_s$.
Generally, SM predictions of $\Delta \Gamma_s / \Gamma_s$
do not exceed 0.2, with a most reliable mean value
$\Delta \Gamma_s / \Gamma_s = 0.12 \pm 0.06$ \cite{duna,lenz,ciu}.
Recently $\Delta \Gamma_s / \Gamma_s = 0.65^{+0.25}_{-0.33} \pm 0.01$ 
\cite{cdfpol} and 
$\Delta \Gamma_s / \Gamma_s = 0.24^{+0.28}_{-0.38}(stat) 
^{+0.03}_{-0.04} (syst)$ \cite{dopol}
have been measured in the $B_s \to J/\Psi \phi$ decays
by the CDF and D0 collaborations, respectively.
While these results are inconclusive due to very large 
statistical uncertainties, the
data samples collected by CDF and D0 are rapidly increasing 
and a much better accuracy can be reached soon.

\subsection{Polarization in {\boldmath $B^0 \to J/\Psi K^{*0}$} and
{\boldmath $B_s \to J/\Psi \phi$} decays}

Here we would like to propose a new model-dependent method to obtain
a $\Delta \Gamma_s / \Gamma_s$ upper limit. This method is based on 
comparing the $CP$-even and 
$CP$-odd components in the $B^0 \to J/\Psi K^{*0}$ and 
$B_s \to J/\Psi \phi$ decays. In the limit of flavor SU(3) symmetry, 
the ratio of the relative components in $B^0 \to J/\Psi K^{*0}$
is expected to be the same as those at proper time $t=0$ in the 
decays $B_s \to J/\Psi \phi$ \cite{dunb}. 
Integrating over corresponding exponential lifetime functions, 
the relative $CP$-even and $CP$-odd components should be corrected
by factors $1/\Gamma_s^{(+)}$ and $1/\Gamma_s^{(-)}$,
respectively. Then, the time integrated $CP$-odd component
in the $B_s \to J/\Psi \phi$ decay can be related to those
in the $B^0 \to J/\Psi K^{*0}$ decay:

\begin{equation}
R_{\perp} (J/\Psi \phi) \approx \frac{ R_{\perp} (J/\Psi K^{*0})}
{1 - ( 1 - R_{\perp} (J/\Psi K^{*0}) \times \frac{\Delta \Gamma_s}{\Gamma_s}}
\approx \frac{0.195}{1 - 0.805 \times \frac{\Delta \Gamma_s}{\Gamma_s}}
\end{equation}
where the Belle collaboration measurement of the $CP$-odd component
$R_{\perp} = |A_{\perp}|^2 = 0.195 \pm 0.012 \pm 0.008$ in the
$B^0 \to J/\Psi K^{*0}$ decay \cite{belpol} is used.
Recently the CDF collaboration has measured
$|A_{\perp}| = 0.354 \pm 0.098 \pm 0.003$ in the
$B_s \to J/\Psi \phi$ decay \cite{cdfpol}, which corresponds
to $R_{\perp} = 0.125 \pm 0.069 \pm 0.002$.
The D0 collaboration has also measured the $CP$-odd component to be
$R_{\perp} = 0.16 \pm 0.10 (stat) \pm 0.02 (syst)$
in the $B_s \to J/\Psi \phi$ decay \cite{dopol}.
Combining the CDF and D0 measurements, the weighted
value of $CP$-odd component
is $R_{\perp}(J/\Psi \phi) = 0.136 \pm 0.057$.

\begin{table*}[h!]
\caption{The experimental measurements and a prediction obtained from
Eq.~4 of the $CP$-odd component in the $B^0 \to J/\Psi K^{*0}$ and 
$B_s \to J/\Psi \phi$ decays.}
\begin{center}
\begin{tabular}{|l|c|c|}
\hline
Decay mode & measurement/ & $CP$-odd  \\
           &  prediction  & component rate \\
\hline
$B^0 \to J/\Psi K^{*0}$ & BELLE & 0.195 $\pm$ 0.012 $\pm$ 0.008 \\
$B_s \to J/\Psi \phi$ & CDF+D0 & 0.136 $\pm$ 0.057 \\
$B_s \to J/\Psi \phi$ & Eq. (4), $\Delta \Gamma_s / \Gamma_s = 0.12$ & 0.216 \\
\hline
\end{tabular}
\end{center}
% \end{ruledtabular}
\vspace{-0.1cm}
\end{table*}

As we can see from Table 1, in contrast to the estimate obtained from Eq.~(4),
$R_{\perp}(J/\Psi \phi)$ is smaller than $R_{\perp}(J/\Psi K^*)$,
although the statistical errors
are too large for definite conclusions. Nonetheless an
upper limit for $\Delta \Gamma_s / \Gamma_s$
can be obtained.
Using the weighted value of $R_{\perp}(J/\Psi \phi)$,
an upper limit $\Delta \Gamma_s / \Gamma_s < 0.08$ at 90$\%$ $CL$ is set
from Eq.~(4), where $CP$-violating effects are assumed 
to be small.
Although additional systematic uncertainties
can increase the upper limit to $\sim$0.11, this upper limit is
smaller than the $\Delta \Gamma_s / \Gamma_s$ values
obtained in direct measurements. Note also
that BSM $CP$-violating effects in
$B_s$ decays can potentially decrease $R_{\perp}(J/\Psi \phi)$ 
and the upper limit will weaken.
Finally, within the Standard Model, the obtained upper limit rules out the 
large values of $\Delta \Gamma_s / \Gamma_s$ reported, 
with large uncertainties, by the CDF and D0 experiments.

\subsection{Comparison of {\boldmath $B^0$} and {\boldmath $B_s$} lifetimes}

Another method estimating of $\Delta \Gamma_s / \Gamma_s$ was
proposed in ref.\cite{hart}. This method is based on the
prediction that 
$\overline{\tau} (B_s) \approx \tau (B^0)$ with
$\pm\,1\%$ accuracy \cite{ben,keum,gab,fra}, where 
$\overline{\tau} (B_s)$ is a mean over long-living and short-living components,
$\overline{\tau} (B_s) = 1 / ( \Gamma_L + \Gamma_S )$.
Assuming a two exponential decay rate for untagged $B_s$ decays,
$\Gamma (B_s^{un}) = A\, {\rm exp}(-\Gamma_L t) + B\, {\rm exp}(-\Gamma_S t)$,
and equal amplitudes ($A = B$) for any flavor-specific $B_s$
lifetime measurement, the mean lifetime measurement in any flavor-specific
decay mode gives a larger value than $\overline{\tau} (B_s)$:
 
\begin{equation}
\overline{\tau}_{\rm fs}  = \ \overline{\tau} (B_s) \ \
\frac{1 + (\Delta \Gamma_s/ 2 \Gamma_s)^2}
{1 - (\Delta \Gamma_s/ 2 \Gamma_s)^2} \ \approx \ \ \overline{\tau} (B_s) \
[\ 1 + 2 (\Delta \Gamma_s/ 2 \Gamma_s)^2\ ]
\end{equation}

As noted in \cite{hart} the excess of the 
flavor-specific $B_s$ lifetime
over the $B^0$ lifetime could provide some estimate of 
the $\Delta \Gamma_s / \Gamma_s$.
However, flavor-specific lifetime measurements give a lifetime
value smaller that the $B^0$ lifetime \cite{ave} (Table 2).
In contrast to expectations, the experimentally measured value is
$\overline{\tau} (B_s) / \tau(B^0) = 0.914 \pm 0.030$ \cite{ave},
that is 2.9$\sigma$ below unity.

\begin{table*}[h!]
\caption{Summary of lifetimes of $B^0$ and $B_s$ mesons \cite{ave}.}
\begin{center}
\begin{tabular}{|l|c|}
\hline
$b$-hadron & Measured lifetime \\
\hline
$B^0$  & 1.527 $\pm$ 0.008 ps \\
$B_s$ ($\to$ flavor specific) & 1.454 $\pm$ 0.040 ps \\
$B_s$ ($\to J/\Psi \phi$) & 1.404 $\pm$ 0.066 ps \\
$B_s$ (1/$\Gamma_s$) & 1.396$^{+0.044}_{-0.046}$ ps \\ 
\hline
\end{tabular}
\end{center}
% \end{ruledtabular}
\vspace{-0.1cm}
\end{table*}

\section{Conclusion}

In conclusion,
the $B_s \to D_s^{(*)+} D_s^{(*)-}$ decay branching fraction
can be measured with 25$\%$ accuracy using a 50$\,$fb$^{-1}$ data
sample collected at an $e^+ e^-$ collider experiment running 
on the $\Upsilon$(5S) resonance. From this measurement,
$\Delta \Gamma^{CP}_s / \Gamma_s$ can be obtained
with 25$\%$ accuracy. Comparing this value with
those obtained in direct $\Delta \Gamma_s / \Gamma_s$
measurements can provide an important
test of the Standard Model.

Two methods can be used to estimate
$\Delta \Gamma_s / \Gamma_s$, the comparison of polarization
in $B^0 \to J/\Psi K^{*0}$ and $B_s \to J/\Psi \phi$ decays
and the comparison of $B^0$ and $B_s$ lifetimes.
In a model-dependent way, we obtain from the polarization method
the upper limit $\Delta \Gamma_s / \Gamma_s < 0.11$ at 90$\%$ $CL$.
This upper limit is below $\Delta \Gamma_s / \Gamma_s$ central
values measured (with large uncertainties) in the Tevatron experiments. 
An even stronger upper limit can be obtained from the lifetime method,
assuming equality of the $B_s$ and $B^0$ lifetimes.

Both polarization and lifetime methods appear to disagree with expectations,
although the experimental uncertainties are still too large for definite
conclusions.
Assuming that theoretical models are correct, low
values of $\Delta \Gamma_s / \Gamma_s$ are favored,
with a only few percent probability to obtain 
$\Delta \Gamma_s / \Gamma_s > 0$.
An excess of $CP$-even decays due to $B_s \to D_s^{(*)+} D_s^{(*)-}$
modes cannot explain these discrepancies.
Additional theoretical studies are required to explain
the experimental data.

The measurements at $e^+ e^-$ collider experiments running
on the $\Upsilon$(5S) resonance can help to shed light on
these discrepancies between experimental measurements of 
$B_s$ decays and theoretical estimates.
The $B_s \to D_s^{(*)+} D_s^{(*)-}$ decay measurements
are very important to understand $B_s$ behaviors.
Potentially, direct lifetime measurements of flavor-specific
and $CP$-specific $B_s$ decays (maybe with a partial $B_s$ reconstruction) 
at the $\Upsilon$(5S) can be also performed.
% to provide important information 
% about $B_s$ mesons.

\end{document}